An Evaluation of Percentile Measures of Citation Impact,

and a Proposal for Making Them Better

Lutz Bornmann*

Richard Williams**

* Division for Science and Innovation Studies

Administrative Headquarters of the Max Planck Society

Hofgartenstr. 8,

80539 Munich, Germany.

Email: bornmann@gv.mpg.de

** Department of Sociology

University of Notre Dame

4058 Jenkins Nanovic Hall

Notre Dame, IN 46556 USA

Email: rwilliam@nd.edu

**Abstract**

Percentiles are statistics pointing to the standing of a paper's citation impact relative to other papers in a given citation distribution. Percentile Ranks (*PR*s) often play an important role in evaluating the impact of scholars, institutions, and lines of study. Because *PR*s are so important for the assessment of scholarly impact, and because citation practices differ greatly across time and fields, various percentile approaches have been proposed to time- and field-normalize citations. Unfortunately, current popular methods often face significant problems in time- and field-normalization, including when papers are assigned to multiple fields or have been published by more than one unit (e.g., researchers or countries). They also face problems for estimating citation counts (*CC*s) for pre-defined *PR*s (e.g., the 90$^{th}$ *PR*). We offer a series of guidelines and procedures that, we argue, address these problems and others and provide a superior means to make the use of percentile methods more accurate and informative. In particular, we introduce two approaches, *CP-IN* and *CP-EX*, that should be preferred in bibliometric studies because they consider the complete citation distribution. Both approaches are based on cumulative frequencies in percentages (*CP*s). The paper further shows how bar graphs and beamplots can present *PR*s in a more meaningful and accurate manner.

# 1 Introduction

Since the 1980s, various methods have been introduced (Tahamtan & Bornmann, 2018) to time- and field-normalize citations. Ioannidis, Boyack, and Wouters (2016) discuss the pros and cons for normalizations of the different methods. In one of the most frequently used methods – the relative citation rate (Schubert & Braun, 1986) – expected values are calculated for every combination of publication year and subject category in databases such as Web of Science (WoS, Clarivate Analytics) or Scopus (Elsevier). These expected values are mean numbers of citations. Then, the citation impact of every focal paper in the combination of publication year and subject category is normalized by dividing the citation counts (*CC*s) of the focal paper by the corresponding expected value (Schubert & Braun, 1986; see also the review of Waltman, 2016).

This approach of generating normalized citation impact values which can be used for cross-time and cross-field comparisons has been frequently criticized, since it is based on arithmetic averages of citations and citations as a rule are skewed distributed. In case of skewed distributions, the arithmetic average should not be used as a measure for the central tendency of the distribution. As an alternative to this normalization approach, various percentile approaches (plotting positions, *PP*s, and percentile ranks, *PR*s) have been proposed. *PP*s are quantiles of an empirical (or theoretical) distribution whereby quantiles are defined as specific cut points partitioning distributions into subsets.

Since a percentile is defined as "a statistic that gives the relative standing of a numerical data point when compared to all other data points in a distribution" (Lavrakas, 2008), *PP*s can be interpreted as percentiles. Suppose various publication sets from a database that each contain all papers published in a specific combination of publication year and subject category. *PR x* is defined then as the *CC* (at or) below which *x*% (e.g., 90%) of the papers in a certain combination of publication year and subject category falls. Two papers from different combinations of subject category and publication year with exactly the same *PR* may have different *CC*s. The advantage of *PR*s (and *PP*s) is that they are not strongly affected by outliers (highly cited papers) and their interpretation is simple and clear: if a focal paper has a *PR* of 90, then 90% of the papers in the publication year and subject category have a citation impact which is (at or) below the impact of the focal paper. This interpretation of *PR*s makes their use in citation analysis attractive, since it clearly shows the position of the paper in the combination of subject category and publication year which can be compared with the position of other papers.

It is not only possible to calculate *PP*s and *PR*s for every single paper in a database such as WoS. It is popular in bibliometrics (Bornmann, 2014) to identify in every combination of publication year and subject category the papers which belong to the 10% most frequently cited papers (Bornmann, de Moya Anegón, & Leydesdorff, 2012; Narin, 1987; Waltman et al., 2012). These are the papers in the citation distribution at (around) or above the 90$^{th}$ *PR*. The number of these papers can be counted for various units (e.g., for a journal, researcher, university, or country). *P*(top 10%) is the number of top-10% papers and *PP*(top 10%) is the proportion of top-10% papers published by a unit.

In recent years, various percentile approaches have been introduced in bibliometrics. In this study, the different approaches are presented, and their advantages and especially disadvantages explained (see section 3). Our discussion will show that widely used percentile measures are problematic in some fairly common situations. In section 4, we offer an optimized *PR* approach that maintains most of the advantages of percentile measures while overcoming or minimizing their most serious weaknesses. In section 5, we pay attention to an appropriate presenting of percentiles such as bar graphs and beamplots that can make the presentation of *PR* results more meaningful and accurate.

## 2    Datasets used

For discussing the different percentile approaches, two datasets are used: (1) a fictitious small dataset including 21 papers with various *CC*s, and (2) all papers published between 2000 and 2005 with the document type "article" and their *CC*s until the end of 2018. The publication and citation data are from the Max Planck Society's in-house database which is based on the WoS. WoS subject categories, which are sets of similar journals, have been used to compute field-normalized citation impact values (*PP*s and *PR*s). The publication set consists of 6,973,937 articles. However, these are more papers than have been published between 2000 and 2005, since papers which have been assigned to more than one subject category have been considered multiple times (for calculating *PP*s and *PR*s in every combination of publication year and subject category). Without multiple counting of papers, the publication set consists of 4,416,554 articles. The articles received between 0 and 67,582 citations until the end of 2018 (median = 14, mean = 31.76).

## 3    Widely used percentile measures and their disadvantages

### 3.1    Counting highly cited papers

Let us start the discussion of the various percentile approaches with the family of *P*(top *x*%) indicators: the number of papers belonging to the *x*% most-frequently cited papers. The use of *P*(top *x*%) in research evaluation might be interpreted as unsatisfying, since the diverse citation impact of a unit's papers (e.g., the papers published by a researcher) is transformed into a binary information: one part of the papers belong to the top *x*% and the other part not. Imagine researcher A has published many papers which are all in the range of *P*(top 11%) and *P*(top 31%) and another researcher B who has published all papers in the range of *P*(top 71%) and *P*(top 91%). Based on these numbers, one can conclude that researcher A has a better performance than researcher B. However, if only the number of *P*(top 10%) are counted, both researchers would receive the same assessment with *P*(top 10%) = 0.

Furthermore, the calculation of *P*(top *x*%) is affected by the problem of citation ties at the threshold for separating the *x*% most frequently cited papers from the rest: suppose five papers with 20 citations, 20 papers with 10 citations, and 75 papers with 1 citations. It is not clear with this citation distribution whether the 20 papers with 10 citations should be assigned to the *P*(top 10%) or the bottom 90%. Although Waltman and Schreiber (2013) found an elegant solution for that problem (leading to a fractional assignment of papers at the threshold to the group of highly cited papers), it leads to data which are no longer binary: the papers at the threshold are counted with a value less than 1. The consequence is, for instance, that the data can no longer be analyzed with logistic regression analyses, although the nature of the indicator (papers belonging to the top *x*% or not) would suggest that this is the appropriate method.

As the example above with researchers A and B reveal, it is desirable to have a percentile solution which is able to reflect the whole range of citation impact received by the papers in a certain publication set. With the integrated impact indicator (*I3*), Leydesdorff and Bornmann (2011, 2012) proposed a solution going beyond the binary classification of impact. Here, the papers in a set (of a journal or a university) are assigned to more than two impact classes based on *PR*s [e.g., six classes; *P*(top 1, 5, 10, 25, 50, 75%)]. The formula for calculating *I3* is

$$I3 = \sum_{i=1}^{y}(x_i * W_i) \qquad (1)$$

whereby *y* is the number of *PR* classes (PCs) *i* considered (e.g., six classes as mentioned above) and $W_i$ denotes the weight of $PC_i$. $x_i$ is the number of papers published by a unit in $PC_i$. As equation 1 suggests, *I3* is basically an instruction for the aggregation of *PC*s. The *PC* including the papers with the most citation impact should receive the maximum weight. The equation can be used very flexible. For example, papers in the highest impact class can be given little more (e.g., six in the case of six classes) or significantly more weight than lowly cited papers. For example, Leydesdorff, Bornmann, and Adams (2019) proposed to weight the number of papers in *P*(top 1%) with 100 and the number of papers in *P*(top 10%) with 10. For the purpose of notifying the number of *PC*s and weights used for calculating *I3* in a study, Leydesdorff et al. (2019) proposed to use the general notation

$$I3(PC_1 - W_1, PC_2 - W_2 \ldots PC_n - W_n) \qquad (2)$$

whereby *PC* is the lower threshold of the *PC*, e.g. 99 in case of *P*(top 1%), and *W* the corresponding weight (e.g., 100). *n* defines the number of classes and weights, respectively. The flexibility in the use of *PC*s and weights might be an advantage of *I3*, since the user can adapt the indicator to certain evaluation tasks (see here Bornmann & Marewski, 2019). The disadvantage of this flexibility is, however, that there is no standardized use of *I3* (and the results may not be comparable). Another problem is that the indicator is still based on classes – *P*(top *x*%) actually is an *I3* indicator which can be expressed with the *I3* notation: *I3*(90 – 1) – and does not consider the complete information of citation impact distributions. It is a decisive disadvantage of *I3* that it can be calculated only on the aggregated level, i.e. for groups of papers. That means, it is not possible to calculate *I3* for all papers included in a database such as the WoS and to use the preprocessed data for citation analyses of various units later.

### 3.2  Plotting positions (*PP*s)

Bornmann, Leydesdorff, and Mutz (2013) discussed several possibilities to calculate *PP*s for receiving time- and field-normalized citation impact values on the single paper level.

They preferred the calculation of *PP*s based on the rule proposed by Hazen (1914) using the formula

$$PP = \frac{i - 0.5}{n} \qquad (3)$$

For calculating the *PP*s, the papers (published in one publication year and subject category) are sorted in decreasing order of *CC*s and ranking positions are assigned, whereby $i$ is the rank of the paper and $n$ is the total number of papers in the set. As the example in Table 1 shows, the formula returns values which are between 0 and 1; for receiving percentages, the values can be multiplied by 100. Ties in citation data do not pose a problem for the calculation, since the corresponding papers simply receive the same (mean) rank and *PP*. *PP*s can be calculated for the papers in all subject categories and publication years in a database such as WoS, whereby one receives comparable time- and field-normalized citation impact values. For example, the results by Bornmann and Marx (2015) reveal favorable results of *PP*s based on the rule proposed by Hazen (1914) compared to other time- and field-normalization methods (e.g., methods based on mean citations, see above).

Table 1. Example set of papers for calculating plotting positions (*PP*s) (21 papers)

| Paper | Citation count | Rank | Hazen | In percent |
|-------|----------------|------|-------|------------|
| A | 20 | 20.5 | 0.95 | 95.24 |
| B | 20 | 20.5 | 0.95 | 95.24 |
| C | 13 | 18.5 | 0.86 | 85.71 |
| D | 13 | 18.5 | 0.86 | 85.71 |
| E | 10 | 17 | 0.79 | 78.57 |
| F | 9 | 16 | 0.74 | 73.81 |
| G | 8 | 14.5 | 0.67 | 66.67 |
| H | 8 | 14.5 | 0.67 | 66.67 |
| I | 7 | 11.5 | 0.52 | 52.38 |
| J | 7 | 11.5 | 0.52 | 52.38 |
| K | 7 | 11.5 | 0.52 | 52.38 |
| L | 7 | 11.5 | 0.52 | 52.38 |
| M | 3 | 9 | 0.40 | 40.48 |
| N | 2 | 8 | 0.36 | 35.71 |
| O | 1 | 6 | 0.26 | 26.19 |

| P | 1 | 6 | 0.26 | 26.19 |
| Q | 1 | 6 | 0.26 | 26.19 |
| R | 0 | 2.5 | 0.10 | 9.52 |
| S | 0 | 2.5 | 0.10 | 9.52 |
| T | 0 | 2.5 | 0.10 | 9.52 |
| U | 0 | 2.5 | 0.10 | 9.52 |

Cox (2005) outlined that it is an important advantage of *PP*s based on the rule proposed by Hazen (1914) that the formula leads to a value of 0.5 (or 50 as percentage) for the single middle value in the citation distribution. However, this is not always the case: since there are several papers with 7 citations in Table 1, a single middle value does not exist. Another problem of the *PP*s concerns their interpretation: a usual definition of a *PR x* is that it represents the *CC* at or below which *x* percent of the papers falls. Four papers in Table 1 have zero citations and a *PP* of 9.52. Thus, one could assume that around 10% of the papers in the table have zero citations; however, there are around 20% of the papers with zero citations. *PP*s have been initially proposed and are calculated for the comparison of two empirical distributions (or an empirical distribution with a theoretical distribution), but not for using them for relative assessments. However, the problem with the interpretation of *PP*s especially concerns small publication sets with only a few papers (fewer than 100 papers). Suppose that there are 100 papers in a set with different *CC*s each. Then, the *PP*s correspond approximately to the percentage of papers at or below the citation impact of the focal paper. For example, the paper with the 10$^{th}$ rank position will have the *PP* of 0.095.

Figure 1 shows a Q-Q plot of quantiles which have been calculated based on the rule proposed by Hazen (1914). Q-Q plots are two-way scatterplots of one variable against another after both variables have been sorted into ascending order (StataCorp., 2017). For all papers published between 2000 and 2005 and assigned to six subject categories in WoS ($n = 6,070$ papers), the quantiles are shown resulting from the first and sixth subject category. In other words, the figure shows the quantiles for two subject categories of one and the same paper.

Since the general trend of the Q-Q plot is on the line $y = x$ which follows the 45° line, the quantiles in both subject categories 1 and 6 are similar. Thus, most of the papers seem to have the same citation impact relative to other papers in the corresponding subject categories.

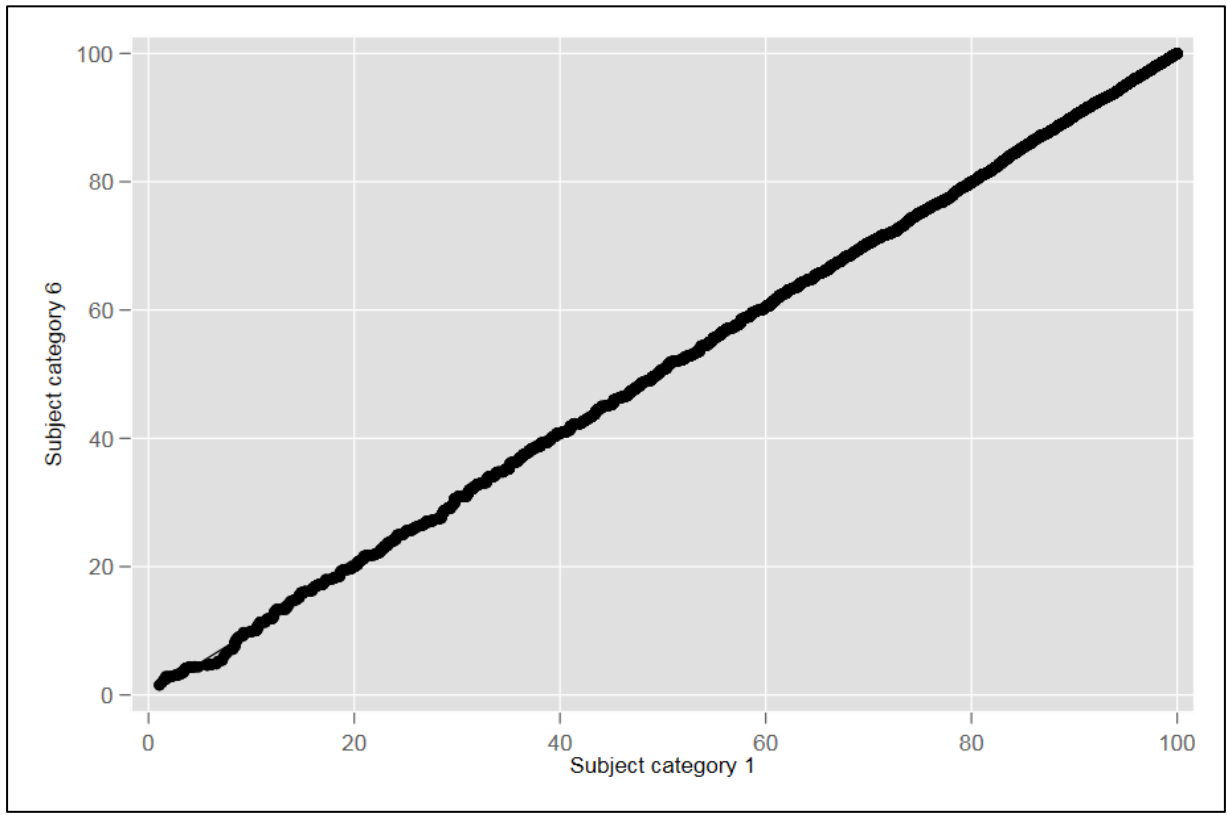

Figure 1. Q–Q plot of quantiles based on the rule proposed by Hazen (1914). For all papers published between 2000 and 2005 and assigned to six subject categories, the quantiles are shown resulting from the first and sixth subject category.

Several other rules have been proposed in the past which can be used instead of Hazen (1914) which lead, however, to similar *PP*s (Cox, 2005, discusses some rules).

### 3.3 Approaches based on size-frequency distributions

In recent years, some other approaches have been proposed for citation analyses which might lead to time- and field-normalized citation impact values. Since these approaches are not used for calculating *PP*s, they are explained in this section based on the size-frequency distribution (Egghe, 2005). This distribution shows the frequencies of papers with certain *CC*s in a set of papers (see the first two columns in Table 2).

The column "InCites" in Table 2 refers to the approach used in the InCites tool which is a citation-based evaluation tool to analyze institutional performance provided by Clarivate Analytics (see https://clarivate.com/webofsciencegroup/solutions/incites). InCites percentiles are cumulative percentages of the size-frequency distribution starting with the percentage of papers with the highest citations. For example, 19.05% of the papers in Table 2 have 13 or more citations; 9.52% of the papers have at least 20 citations. Clarivate Analytics defines InCites percentiles as the "percentage of papers at each level of citation, i.e., the percentage of papers cited more often than the paper of interest" (see https://clarivate.libguides.com/incites_ba/alpha-indicators). Other than the Hazen approach, the InCites approach provides normalized values which can be interpreted as an exact percentage of papers. It is another advantage of the InCites approach that the number (percentage) of top $x$% papers in a paper set can be quickly identified. The problem, however, with the InCites approach is that one does not immediately know, how good a focal paper is compared to the other papers in the set. Only the subtraction from 100, i.e. $(100 - x)$, reveals the percentage of papers performing worse than the focal paper.

Table 2. Example set of papers for calculating various time- and field-normalized values based on size-frequency distributions (based on the same 21 papers as in Table 1)

| Citation count | Number of papers | Rank $k$ | InCites | Rank $i$ | *P100* | Rank $j$ | *P100'* |
|---|---|---|---|---|---|---|---|
| 0 | 4 | 21 | 100.00 | 0 | 0.00 | 0 | 0.00 |
| 1 | 3 | 17 | 80.95 | 1 | 11.11 | 4 | 21.05 |
| 2 | 1 | 14 | 66.67 | 2 | 22.22 | 7 | 36.84 |
| 3 | 1 | 13 | 61.90 | 3 | 33.33 | 8 | 42.11 |
| 7 | 4 | 12 | 57.14 | 4 | 44.44 | 9 | 47.37 |
| 8 | 2 | 8 | 38.10 | 5 | 55.56 | 13 | 68.42 |
| 9 | 1 | 6 | 28.57 | 6 | 66.67 | 15 | 78.95 |
| 10 | 1 | 5 | 23.81 | 7 | 77.78 | 16 | 84.21 |
| 13 | 2 | 4 | 19.05 | 8 | 88.89 | 17 | 89.47 |

| 20 | 2 | 2 | 9.52 | 9 | 100.00 | 19 | 100.00 |

Bornmann, Leydesdorff, and Wang (2013) introduced the *P100* approach which does not use the size-frequency distribution as the InCites approach, but the distribution of unique citation values (see Table 2). Thus, the frequencies of papers with certain *CC*s are not considered. The formula is

$$P100 = \frac{i}{i_{max}} * 100 \qquad (4)$$

whereby *i* is the rank of the citation in the distribution of unique citation values (see Table 2). According to Bornmann and Mutz (2014) *P100*, however, "has undesirable properties which should be avoided … [for example,] the scale value of a paper can increase as a result of the fact that another paper receives an additional citation" (p. 1940). Another problem with this approach is similar to that of the Hazen approach: a *P100* value does not refer to the *CC* at or below which *x* percent of the papers in the combination of publication year and subject category falls. Thus, Bornmann and Mutz (2014) introduced *P100'* as an alternative to *P100* which is calculated using the formula

$$P100' = \frac{j}{j_{max}} * 100 \qquad (5)$$

whereby *j* are ranks based on the size-frequency distribution (see Table 2). It is a decisive advantage of *P100'* that the indicator always has the maximum value 100 and the minimum value 0. However, it remains the problem (as with the *P100* and *PP* indicators) that it cannot be interpreted properly: *P100'* = 21.05 does not mean that 21.05% of the papers in the publication set are below 1 citation (or equal to that *CC*).

# 4 Cumulative frequencies in percentages as an optimized percentile approach

The problem of proper interpretations of percentiles can be solved by calculating cumulative frequencies in percentages (*CP*s) as demonstrated in Table 3 (and by the InCites approach). The table includes two variants: *CP-IN* is the cumulative percentage of the size-frequency distribution of papers. For *CP-EX*, the first possible percentage is set at 0 rather than considering its actual cumulative percentage. Then, the calculation of the cumulative percentage starts with the percentage of the lowest *CC*. In this way, *CP-EX* reveals exactly the percentage of papers with lower citation impact: for example, *CP-EX* = 90.48 means that 90.48% of the papers in the set received a citation impact which **is below** 20 citations; 19.05% of the papers received less than one citation. *CP-IN* has a slightly other interpretation: 90.48 means that 90.48% of the papers in the set received a citation impact which **is at or below** 13 citations; 19.05% of the papers received zero citations.

Table 3. Cumulative percentages including (*CP-IN*) or excluding (*CP-EX*) the number of papers in the row (based on the same 21 papers as in the previous tables)

| Citation count | Number of papers | Percent (including papers in row) | Cumulative percentage (*CP-IN*) | Percent (excluding papers in row) | Cumulative percentage (*CP-EX*) |
|---|---|---|---|---|---|
| 0 | 4 | 19.05 | 19.05 | 0.00 | 0.00 |
| 1 | 3 | 14.29 | 33.33 | 19.05 | 19.05 |
| 2 | 1 | 4.76 | 38.10 | 14.29 | 33.33 |
| 3 | 1 | 4.76 | 42.86 | 4.76 | 38.10 |
| 7 | 4 | 19.05 | 61.90 | 4.76 | 42.86 |
| 8 | 2 | 9.52 | 71.43 | 19.05 | 61.90 |
| 9 | 1 | 4.76 | 76.19 | 9.52 | 71.43 |
| 10 | 1 | 4.76 | 80.95 | 4.76 | 76.19 |
| 13 | 2 | 9.52 | 90.48 | 4.76 | 80.95 |
| 20 | 2 | 9.52 | 100.00 | 9.52 | 90.48 |
| Total | 21 | 100.00 | | | |

*CP-IN* and *CP-EX* have been calculated for all articles in the dataset of this study published between 2000 and 2005. Figure 2 (upper left side) shows the distribution of *CC*s for the six years using boxplots. It is clearly visible that the distributions are very skewed and characterized by outliers (by a few highly cited articles).

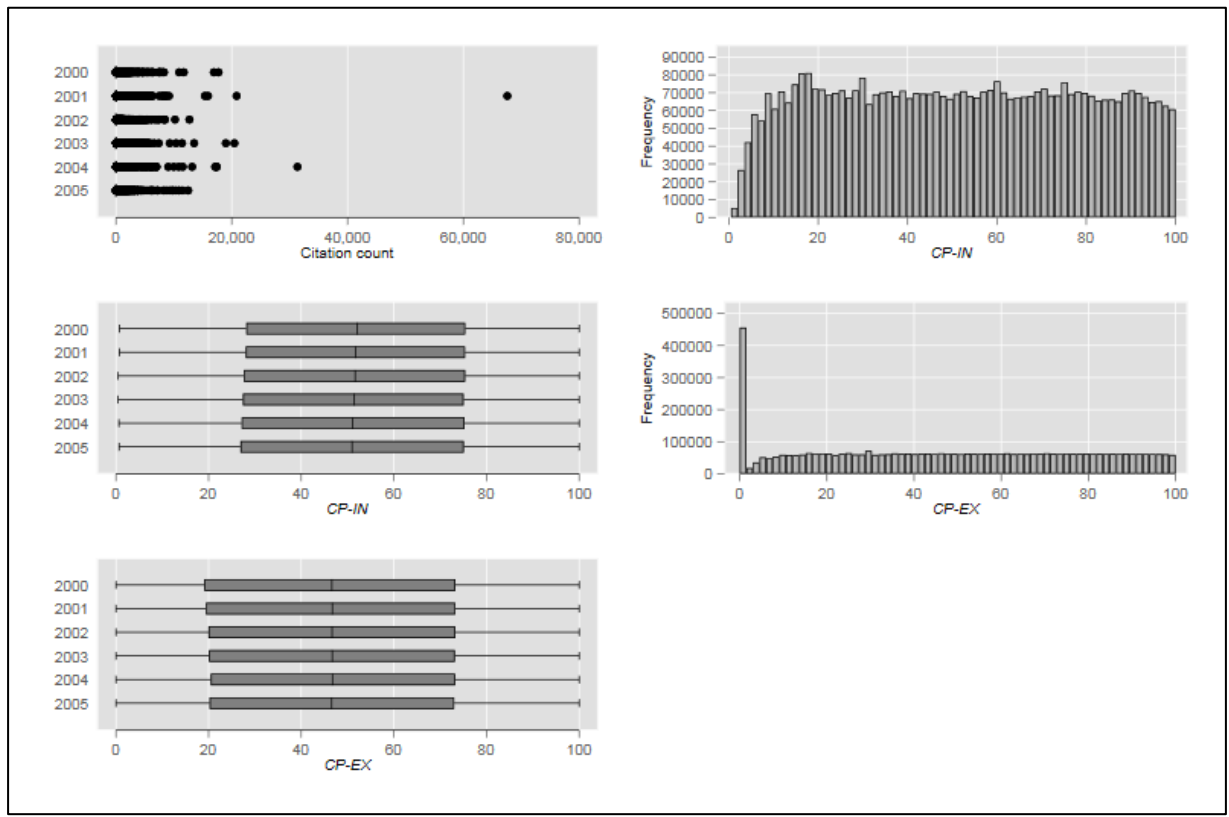

Figure 2. Boxplots showing the distributions of citation counts (*CC*s), *CP-IN*s, and *CP-EX*s (left side) as well as histograms of *CP-IN*s and *CP-EX*s (right side) for articles published between 2000 and 2005

The boxplots of *CP-IN*s and *CP-EX*s in Figure 2 (left side) show that the distributions changed compared to *CC*s: they are not characterized by outliers and are in the range between about 0 and 100. The median *CP-IN* is between 51 (in 2005) and 52 (in 2000); the median *CP-EX* is approximately 46.6 (in all publication years). Figure 2 (right side) shows histograms of *CP-IN*s and *CP-EX*s for articles published between 2000 and 2005. Both figures reveal that the *CP-IN*s and *CP-EX*s are uniformly distributed between values from around 10 to 100. *CP-IN*s and *CP-EX*s below 10 are significantly less frequent than the other *PR* values. The reason is that low values are underrepresented, because the subject categories usually include a lot of papers with zero or only a few citations. The exception, however, is the *PR* zero which is the most frequent value for *CP-EX*. Since the lowest citation impact value in every subject category receives the *PR* zero, this result can be expected for *CP-EX*.

*PR*s such as *CP-IN* and *CP-EX* can be calculated for all papers in every combination of subject category and publication year in databases such as WoS. Then, these papers receive time- and field-normalized citation impact values which can be used for various cross-time and cross-field comparisons (e.g., for the comparison of universities or countries). For these and other practical uses of *PR*s (*CP-IN* and *CP-EX*), however, three problems have to be solved.

**4.1    Solution to the problem of missing comparability**

The first problem is that the *PR*s which have been calculated for different combinations of publication years and subject categories are frequently not comparable: certain *PR*s cannot be calculated based on the data of a certain combination. As the results in Table 2 demonstrate, we know the *CC*s for the *PR*s 90.48 and 42.86, but we do not know the *CC*s for the *PR*s 90 and 50. In other tables, the *CC*s for other *PR*s might be available (e.g., for 82.34 and 23.45). However, for comparing the impact differences between two subject categories, it is necessary to know these *CC*s for predefined *PR*s (e.g., 90 and 50). In this section, we present two approaches which can be used to solve this problem of comparability. The approaches cannot only be used to estimate *CC*s from pre-defined *PR*s, but also to estimate *PR*s from pre-defined *CC*s.

4.1.1    Point-estimates of hypothetical intervals

The approach by Barrett (2003) requires that one takes into account the lower and upper bounds for every *CC* in a citation distribution assuming that "each exact integer score is actually the middle score of an interval extending 0.5 either side" (p. 6). Furthermore, the observable list of *CC*s between the minimum and maximum value in a publication set is complemented by the missing *CC*s (with zero numbers of papers).

Table 4 includes *CP-IN* and shows the same dataset as in the previous tables but considers the lower and upper bounds for each *CC*. Furthermore, the missing *CC*s (between two observable *CC*s) with corresponding zero numbers of papers are added. Thus, a complete list of *CC*s beginning with the minimum and ending with the maximum *CC*s from the initial publication set with the corresponding numbers of papers is available now. In Table 4, the observed *CC*s from the previous tables are expanded to citation-intervals with equal sizes (i.e., 1 citation).

One potential problem with any system using percentiles is measurement error. Barrett (2003) implicitly refers to a measurement error concept with the concept of intervals. Every unit of measurement can have some kind of random variability. For example, with the removal or inclusion of journals in a citation index, citations of papers in the database can change again and again. One can use smaller or wider intervals for this uncertainty. However, intervals have to be set at some size, and any choice is going to be somewhat arbitrary. When in doubt, we think it is best to use common practices, which in this case is using intervals that result in 1 (+/-0.5). Anyone using our methods could, of course, use different-sized intervals if they so wished.

Table 4. Expanding the set of papers for calculating "theoretical" citation counts (*CC*s) for certain *PR*s (*CP-IN*, based on the same 21 papers as in the previous tables)

| Citation count | Lower and upper bounds | Number of papers | Cumulative frequencies | Percent | *CP-IN* |
|---|---|---|---|---|---|
| 0 | [-0.5 to 0.5[ | 4 | 4 | 19.05 | 19.05 |
| 1 | [0.5 to 1.5[ | 3 | 7 | 14.29 | 33.33 |
| 2 | [1.5 to 2.5[ | 1 | 8 | 4.76 | 38.10 |
| 3 | [2.5 to 3.5[ | 1 | 9 | 4.76 | 42.86 |
| 4 | [3.5 to 4.5[ | 0 | 9 | 0.00 | 42.86 |
| 5 | [4.5 to 5.5[ | 0 | 9 | 0.00 | 42.86 |
| 6 | [5.5 to 6.5[ | 0 | 9 | 0.00 | 42.86 |
| 7 | [6.5 to 7.5[ | 4 | 13 | 19.05 | 61.90 |

| 8  | [7.5 to 8.5[    | 2 | 15 | 9.52 | 71.43  |
|----|-----------------|---|----|------|--------|
| 9  | [8.5 to 9.5[    | 1 | 16 | 4.76 | 76.19  |
| 10 | [9.5 to 10.5[   | 1 | 17 | 4.76 | 80.95  |
| 11 | [10.5 to 11.5[  | 0 | 17 | 0.00 | 80.95  |
| 12 | [11.5 to 12.5[  | 0 | 17 | 0.00 | 80.95  |
| 13 | [12.5 to 13.5[  | 2 | 19 | 9.52 | 90.48  |
| 14 | [13.5 to 14.5[  | 0 | 19 | 0.00 | 90.48  |
| 15 | [14.5 to 15.5[  | 0 | 19 | 0.00 | 90.48  |
| 16 | [15.5 to 16.5[  | 0 | 19 | 0.00 | 90.48  |
| 17 | [16.5 to 17.5[  | 0 | 19 | 0.00 | 90.48  |
| 18 | [17.5 to 18.5[  | 0 | 19 | 0.00 | 90.48  |
| 19 | [18.5 to 19.5[  | 0 | 19 | 0.00 | 90.48  |
| 20 | [19.5 to 20.5[  | 2 | 21 | 9.52 | 100.00 |

In the table, each exact *CC* is the middle value of an interval extending 0.5 either side. Using the formula

$$CC_i = lb_i + \left(\frac{n * p - cf_{i-1}}{f_i}\right) * w \qquad (6)$$

the *CC* for a certain *PR i* can be calculated, where

$CC_i$ = *CC* for the $i^{th}$ *PR*

$lb_i$ = the exact lower bound of the interval containing the *CC* for the $i^{th}$ *PR*

$n$ = the total number of papers in the publication set

$p$ = the proportion corresponding to the desired *PR* (between 0 and 1, instead of 0 and 100)

$cf_{i-1}$ = the cumulative frequency of papers in the interval containing the $i\text{-}1^{th}$ *PR*

$f_i$ = the frequency of papers in the interval containing the $i^{th}$ *PR*

$w$ = the width of the class interval

The resulting *CC*s for pre-specified *PR*s are "estimates of hypothetical real-valued continuous numbers" (Barrett, 2003, p. 9). These estimates can be calculated for *PR*s which are between the minimum and maximum *CP-IN* in Table 4 (i.e., 19.05 and 100). Thus, the *CC* for the $4^{th}$ *PR* cannot be estimated.

Using the frequency distribution in Table 4, the *CC*s for the $90^{th}$, $75^{th}$, and $50^{th}$ *PR*s are exemplarily calculated in the following. Let us start with the $90^{th}$ *PR*. Looking at *CP-IN* in the table, one can see that the 90th *PR* is positioned between 12 citations (*CP-IN* = 80.95) and 13 citations (*CP-IN* = 90.48). We can expect that the 90th *PR* is in the interval between 12.5 and 13.5 citations; 13 citations refer to the *PR* 90.48 which is close to 90. The lower bound of the interval is 12.5, i.e. *lb* = 12.5. We are interested in the 90th *PR*, thus *p* = 0.9, and there are 21 papers in the publication set (*n* = 21). The width of the class interval is 1 citation (*w* = 1). The frequency and cumulative frequency of papers containing the *PR* is *f* = 2 and *cf* = 17. Filling these values in the formula leads to an estimate of 13.45 citations. Since 13 citations correspond to the *PR* 90.48 and the upper bound of the interval containing this *PR* is (around) 13.5, 13.45 citations seems to be a realistic value.

The calculation of the estimated *CC* for the 75th *PR* is similar. This *PR* is between 8 and 9 citations in Table 4. Thus, *lb* = 8.5. The interval, in which 8.5 is the lower bound, refers to the *PR* 76.19. The other values for the formula are *n* = 21, *f* = 1, *cf* = 15, *w* = 1, and *p* = 0.75. The estimated result for the 75th *PR* is 9.25 which is close to the upper bound of the interval (around 9.5). This upper bound is from the *PR* 76.19 which is somewhat higher than 75. The last example is the 50th *PR* which is between 6 and 7 citations. Thus, the values for the formula are *lb* = 6.5, *n* = 21, *f* = 4, *cf* = 9, *w* = 1, and *p* = 0.5. The estimated *CC* is 6.875 – a realistic value with around 7.5 as upper bound for the *PR*s 61.9 and 5.5 as lower bound for the *PR* 42.86.

Using the formula

$$PR_x = \left[\frac{\left(cf_i + \left(\frac{x - lb_i}{w}\right) * f_i\right)}{n}\right] * 100 \qquad (7)$$

and the values as specified above, the corresponding *PR*s (*PR*$_x$) can be calculated. The only new parameter is *x*. This is the *CC* for which the *PR* is calculated. Filling in *lb* = 6.5, *n* = 21, *f* = 4, *cf* = 9, *w* = 1, and *x* = 6.875 (for the corresponding *PR i* in case of *lb*, *f*, and *cf*), we receive the PR 50. The formula can be used to determine (estimate) the different PRs for the same CC in different combinations of publication year and subject category. Thus, differences in CCs between two combinations of publication years and subject categories can be made visible based on estimations: what is the PR for the same CC in two different combinations of publication years and subject categories?

The same calculations as with CP-IN can be done with CP-EX but with slightly different formulas. Table 5 shows the same dataset as in Table 4 but it includes CP-EX instead of CP-IN.

Table 5. Expanding the set of papers for calculating "theoretical" citation counts (CCs) for certain PRs (CP-EX, based on the same 21 papers as in the previous tables)

| Citation count | Lower and upper bounds | Number of papers | Cumulative frequencies | Percent | CP-EX |
|---|---|---|---|---|---|
| 0 | [-0.5 to 0.5[ | 4 | 0 | 19.05 | 0.00 |
| 1 | [0.5 to 1.5[ | 3 | 4 | 14.29 | 19.05 |
| 2 | [1.5 to 2.5[ | 1 | 7 | 4.76 | 33.33 |
| 3 | [2.5 to 3.5[ | 1 | 8 | 4.76 | 38.10 |
| 4 | [3.5 to 4.5[ | 0 | 9 | 0.00 | 42.86 |
| 5 | [4.5 to 5.5[ | 0 | 9 | 0.00 | 42.86 |
| 6 | [5.5 to 6.5[ | 0 | 9 | 0.00 | 42.86 |
| 7 | [6.5 to 7.5[ | 4 | 9 | 19.05 | 42.86 |
| 8 | [7.5 to 8.5[ | 2 | 13 | 9.52 | 61.90 |
| 9 | [8.5 to 9.5[ | 1 | 15 | 4.76 | 71.43 |
| 10 | [9.5 to 10.5[ | 1 | 16 | 4.76 | 76.19 |
| 11 | [10.5 to 11.5[ | 0 | 17 | 0.00 | 80.95 |
| 12 | [11.5 to 12.5[ | 0 | 17 | 0.00 | 80.95 |
| 13 | [12.5 to 13.5[ | 2 | 17 | 9.52 | 80.95 |
| 14 | [13.5 to 14.5[ | 0 | 19 | 0.00 | 90.48 |
| 15 | [14.5 to 15.5[ | 0 | 19 | 0.00 | 90.48 |
| 16 | [15.5 to 16.5[ | 0 | 19 | 0.00 | 90.48 |
| 17 | [16.5 to 17.5[ | 0 | 19 | 0.00 | 90.48 |
| 18 | [17.5 to 18.5[ | 0 | 19 | 0.00 | 90.48 |
| 19 | [18.5 to 19.5[ | 0 | 19 | 0.00 | 90.48 |
| 20 | [19.5 to 20.5[ | 2 | 19 | 9.52 | 90.48 |

The formula for calculating the CC for PR i is

$$CC_i = lb_i + 1 + \left(\frac{n * p - cf_i}{f_i}\right) * w \qquad (8)$$

where

$CC_i$ = CC for the $i^{th}$ PR

$lb_i$ = the exact lower bound of the selected interval containing the CC for the $i^{th}$ PR

$n$ = the total number of papers in the publication set

$p$ = the proportion corresponding to the desired PR (between 0 and 1, instead of 0 and 100)

$cf_i$ = the cumulative frequency of papers in the $i^{th}$ PR

$f_i$ = the frequency of papers in the selected interval

$w$ = the width of the class interval

Based on Table 5, the CCs for the 90$^{th}$ PR is exemplarily calculated. This PR is positioned between 13 citations (CP-EX = 80.95) and 14 citations (CP-EX = 90.48). We select the interval in the table with the exact upper bound value (around) 13.5. The lower bound of the interval is 12.5, i.e. lb = 12.5. The other values are $p$ = 0.9, $n$ = 21, $w$ = 1, $f$ = 2, and $cf$ = 17. Filling these values in the formula lead to an estimate of 14.45 citations. Since 14 citations corresponds to the PR 90.48, which is very close to the 90$^{th}$ PR, and the upper bound of the interval containing this PR is around 14.5, 14.45 citations seems to be reasonable.

Using the formula

$$PR_x = \left[\frac{\left(cf_i + \left(\frac{x - lb_i - 1}{w}\right) * f_i\right)}{n}\right] * 100 \qquad (9)$$

and the values from above, the corresponding PR ($PR_x$) can be calculated. Filling in $lb$ = 12.5, $n$ = 21, $f$ = 2, $cf$ = 17, $w$ = 1, and $x$ = 14.45 in the formula (for the corresponding PR $i$ in case of $lb$, $f$, and $cf$), we receive the PR of 90.

4.1.2 Linear interpolation

The second approach of estimating certain CCs or PRs is linear interpolation. If one is interested in the CC for a specific PR of the (non-existent) paper i (e.g., the exact CC for the 90th percentile) and the values cannot be derived from the citation and percentile data of a certain publication year and subject category combination, linear interpolation can be applied (as an alternative to the approach in the previous section). For a certain $PR_i$ (e.g., 90%), $CC_i$ can be estimated based on adjacent values with the equation

$$CC_i = CC_{i-1} + (PR_i - PR_{i-1})\frac{CC_{i+1} - CC_{i-1}}{PR_{i+1} - PR_{i-1}} \qquad (10)$$

Thus, $CC_i$ is estimated based on two observed combinations of existent paper values: ($CC_{i-1}$; $PR_{i-1}$) and ($CC_{i+1}$; $PR_{i+1}$). Using a similar formula, it is also possible to estimate $PR_i$ based on the same observed combinations of values

$$PR_i = PR_{i-1} + (CC_i - CC_{i-1})\frac{PR_{i+1} - PR_{i-1}}{CC_{i+1} - CC_{i-1}} \qquad (11)$$

Figure 3 shows a scatterplot of CCs and CP-EXs that are derived from Table 3. Two data points are signed with a black cross as examples for illustrating the linear interpolation approach: at the one point, we are interest in $PR_i$ for 5 citations and at the other, we are interested in $CC_i$ for a PR of 80%. For both data points, each adjacent combination ($CC_{i-1}$; $PR_{i-1}$) and ($CC_{i+1}$; $PR_{i+1}$) is indicated in the figure. Filling these values in equations 10 and 11, result in 12.4 citations for the 80th PR and a PR of 40.48% for 5 citations. These are plausible estimated values given the observed values in the figure.

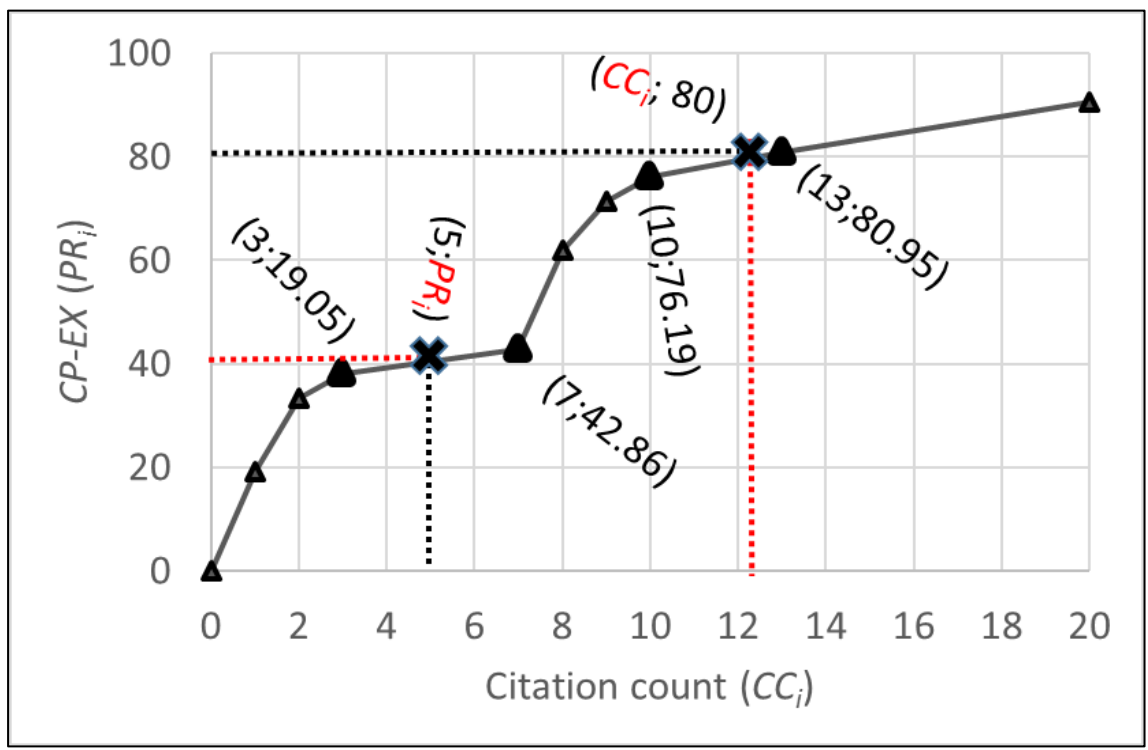

Figure 3. Scatterplot of citation counts ($CC_i$) and *CP-EX*s ($PR_i$). The presented values are from Table 3.

### 4.2 Solution to the problem of multiple field-specific assignments of papers

Using two different approaches, the first problem – certain *PR*s cannot be calculated based on the citation data of a certain combination of publication year and subject category – can be solved. Let us go on with the second problem. Many papers in the WoS and Scopus databases (and in other databases) are assigned to more than one subject category. Between 2000 and 2005, the 4,416,554 articles in the dataset of this study are assigned to up to six subject categories (which leads to 6,973,937 articles including multiple occurrences).

Table 6. Frequency and percentage of articles with different numbers of subject categories

| Number of subject categories | Frequency | Percent | Cumulative frequencies |
|---|---|---|---|
| 1 | 2,583,330 | 58.49 | 58.49 |
| 2 | 1,257,957 | 28.48 | 86.97 |
| 3 | 456,562 | 10.34 | 97.31 |
| 4 | 94,588 | 2.14 | 99.45 |
| 5 | 18,047 | 0.41 | 99.86 |
| 6 | 6,070 | 0.14 | 100.00 |

| Total | 4,416,554 | 100.00 | |

Table 6 shows that 58.49% of the articles are assigned to one subject category. For around 40% of the articles, more than one *PR* is calculated, and it is not clear how these *PR*s can be aggregated into one value for a single article. For the InCites tool (see above), the minimum value is used: "the category in which the percentile value is closest to zero is used, i.e. the best performing value" (see https://clarivate.libguides.com/incites_ba/understanding-indicators). This approach, however, leads to an overestimation of performance if papers are assigned to more than one subject category and the resulting *PR*s are (very) different. For example, 1,833,224 articles in the dataset of this study are assigned to at least two subject categories. The minimum of the differences between two *PR*s of an article is 0, the maximum is 82.9; the mean difference is 7.84 and the median is 5.77.

Another solution for the aggregation into one value could be the median of the *PR*s. A challenge to this calculation is that subject categories have different numbers of papers. In the dataset of this study, there are 1,528 different combinations of subject category and publication year, with a minimum number of papers of 1 and a maximum number of 43,456 (median = 4,564.1, mean = 7,394.94). We can assume that the same *PR* has a higher value in a combination of publication year and subject category with many papers than in a combination with only a few papers. Suppose a paper is assigned to two subject categories with 10 and 3,000 papers. To be at the 50$^{th}$ *PR* in these subject categories would have very different meanings: in one case, the paper would be better than five other papers and in the other case, it would be better than 1,500 other papers. A good estimation of an aggregated *PR* might be to find the median of *PR*s where each subject category's *PR* is reflected several times proportional to the number of papers in that subject category. Thus, the solution might be the *PR* which is weighted by the number of papers in the subject categories. This weighted (w) *PR* can be calculated using the formula

$$wPR = \frac{(PR_{SC1} * n_{SC1}) + (PR_{SC2} * n_{SC2}) + \cdots + (PR_{SCx} * n_{SCx})}{n_{SC1} + n_{SC2} + \cdots + n_{SCx}} \quad (12)$$

whereby $PR_{SCx}$ is the $PR$ of $SC_x$ (i.e., the $SC$ to which a paper is assigned) and $n$ is the total number of papers in these subject categories.

Table 7. Size-frequency distribution of papers in subject category 1 with two focal papers (A and B)

| Selected focal paper | Citation count | Number of papers | Rank $i$ | $CP$-$EX$ |
|---|---|---|---|---|
| Paper A | 100 | 1 | 67 | 98.53 |
| | 98 | 3 | 64 | 94.12 |
| | 90 | 1 | 63 | 92.65 |
| | 88 | 1 | 62 | 91.18 |
| | 40 | 1 | 61 | 89.71 |
| Paper B | 20 | 5 | 56 | 82.35 |
| | 8 | 7 | 49 | 72.06 |
| | 7 | 5 | 44 | 64.71 |
| | 6 | 9 | 35 | 51.47 |
| | 4 | 13 | 22 | 32.35 |
| | 1 | 22 | 0 | 0.00 |
| Total | | 68 | | |

Table 8. Size-frequency distribution of papers in subject category 2 with two focal papers (A and B)

| Selected focal paper | Citation count | Number of papers | Rank $i$ | $CP$-$EX$ |
|---|---|---|---|---|
| | 1100 | 1 | 410 | 99.76 |
| | 980 | 1 | 409 | 99.51 |
| | 465 | 3 | 406 | 98.78 |
| | 200 | 5 | 401 | 97.57 |
| | 145 | 7 | 394 | 95.86 |
| | 120 | 19 | 375 | 91.24 |
| Paper A | 100 | 2 | 373 | 90.75 |
| | 90 | 4 | 369 | 89.78 |
| | 67 | 6 | 363 | 88.32 |
| | 55 | 5 | 358 | 87.10 |
| | 34 | 8 | 350 | 85.16 |
| | 34 | 1 | 349 | 84.91 |
| | 23 | 4 | 345 | 83.94 |
| | 23 | 5 | 340 | 82.73 |
| | 21 | 12 | 328 | 79.81 |
| Paper B | 20 | 13 | 315 | 76.64 |
| | 11 | 15 | 300 | 72.99 |
| | 10 | 22 | 278 | 67.64 |
| | 10 | 34 | 244 | 59.37 |
| | 1 | 45 | 199 | 48.42 |
| | 1 | 55 | 144 | 35.04 |
| | 0 | 67 | 77 | 18.73 |
| | 0 | 77 | 0 | 0.00 |
| Total | | | 411 | |

Table 7 and Table 8 show the size-frequency distributions (*CP-EX*) of two different subject categories. Two papers A and B are assigned to both subject categories. For example, paper A has 100 citations which means that the paper has *CP-EX* = 98.53 in subject category 1 and *CP-EX* = 90.75 in subject category 2. The calculation of a mean *PR* with (98.53 + 90.75) / 2 = 94.64 gives the *PR* from subject category 1 too much weight, since this subject category has significantly fewer papers than subject category 2. Thus, the *wPR* is 91.86 [((98.53 * 68) + (90.75 * 411)) / (68 + 411)] which is lower than the unweighted mean *PR*. The *wPR* of paper B is 77.45 [((82.35 * 68) + (76.64 * 411)) / (68 + 411)] which is also lower than the unweighted mean *PR* with 79.5.

## 4.3   Solution to the aggregation problem of percentiles

The third problem which must be solved with the use of *PR*s is their aggregation if an 'average' *PR* is desired for a certain publication set (e.g., of a researcher, university or country). Suppose one is interested in an 'average' *PR* of paper A (*CP-EX* = 91.86) and paper B (*CP-EX* = 77.45) from Table 7 and Table 8, since these are two papers from a certain unit. In these cases, where the aggregation of *PR*s is desired for a unit, the mean weighted (*mw*) *PR* can be calculated using the formula

$$mwPR = \left(\frac{wPR_1 + wPR_2 + \cdots + wPR_y}{y}\right) \qquad (13)$$

whereby $wPR_1$ to $wPR_y$ are the sums of the *wPR*s for paper 1 to paper *y* published by the unit which are divided by the number of papers published by the unit (*y*). Thus, for the two papers A and B, the *mwPR* is 84.65 [(91.86 + 77.45) / 2].

The formula

$$mwPR(F) = \left(\frac{(wPR_1 * FR_1) + (wPR_2 * FR_2) + \cdots + (wPR_y * FR_y)}{\sum_{i=1}^{y} FR_i}\right) \qquad (14)$$

extends the calculation by another weight besides the number of subject categories: if papers have been published by more than one unit (e.g., researchers, institutions or countries), the papers should be fractionally assigned to these units (see Waltman & van Eck, 2015; Waltman, van Eck, van Leeuwen, Visser, & van Raan, 2011). For example, if a paper was published by authors from two countries, the paper is weighted by 0.5 (Gauffriau, Larsen, Maye, Roulin-Perriard, & von Ins, 2008, provides an overview of various counting methods). The fractional assignment (weighting) is included by the notation $FR_i$ for paper $i = 1$ to paper *y*.

## 5 Presenting cumulative frequencies in percentages

The publication set used in this study consists of 4,416,554 articles published between 2000 and 2005. In this section, based on this dataset, some graphs are presented how *PR*s can be presented for enabling meaningful interpretations of empirical findings. Authors' country information has been added to the articles in the dataset of this study to enable exemplary analyses and comparisons of units. Instead of countries, any other units could have been used (e.g., researchers or institutions).

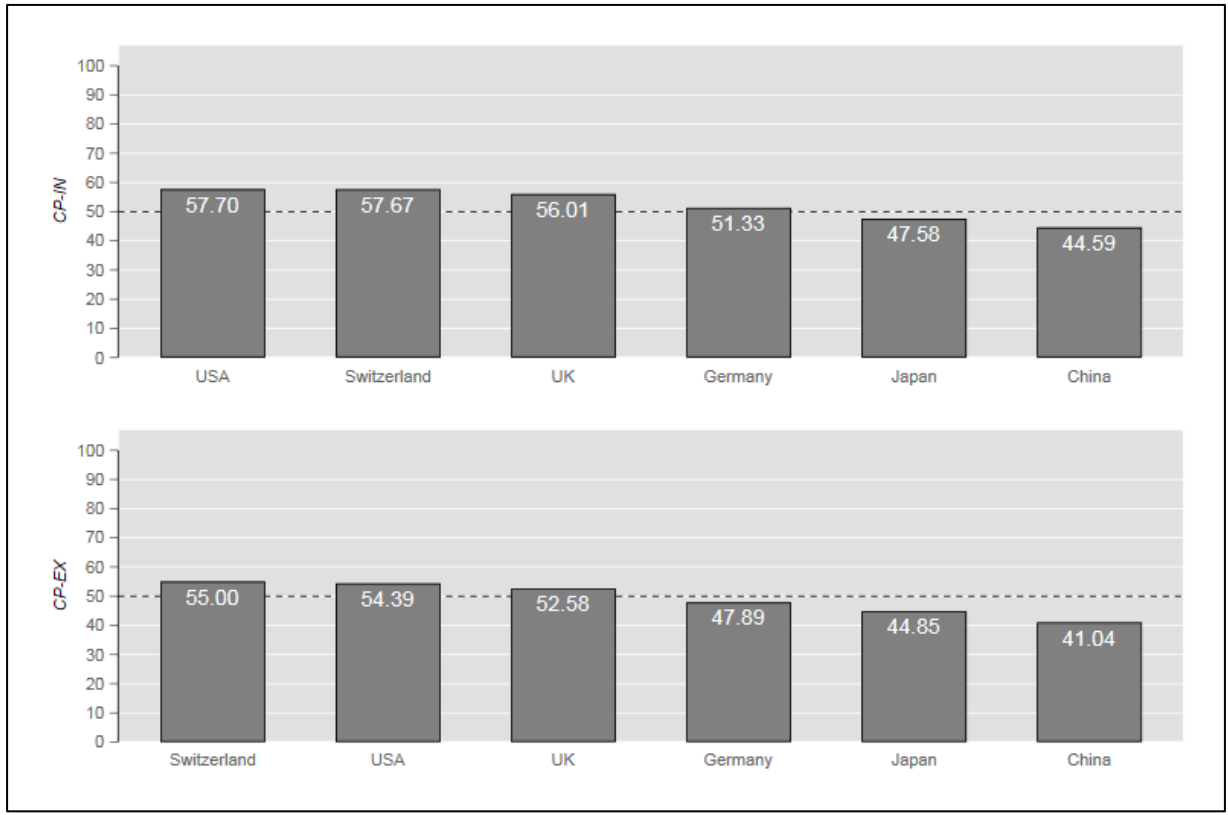

Figure 4. Mean weighted (*mw*) *PR(F)*s (*CP-IN* and *CP-EX*) for articles published between 2000 and 2005 by six countries: USA (*n* = 1,395,809), Switzerland (*n* = 73,344), UK (*n* = 370,227), Germany (*n* = 343,571), Japan (*n* = 372,581), and China (*n* = 255,379). Because of large sample sizes the standard errors of the estimates are too small for presentation in the figure.

The first proposal for presenting *PR*s is to show *mwPR(F)*s as bar graphs. Using the formula in section 4.3, the *mwPR(F)* has been calculated for some countries (by considering two weights: number of papers in a subject category and number of countries). The results are shown in Figure 4: the USA and Switzerland are the best performing countries in this group with *mwPR(F)*s of 57.7 (*CP-IN*) and 54.39 (*CP-EX*) for the USA and 57.67 (*CP-IN*) and 55 (*CP-EX*) for Switzerland. Thus, the articles from these countries performed (equal to or) better than about 58% (55%) of the articles published in the same publication year and subject category. China achieved *mwPR(F)*s of 44.59 (*CP-IN*) and 41.04 (*CP-EX*) which is below the expected value for an 'average' citation level (50). Thus, China's performance is better than around 41% of the articles published in the same publication year and subject category.

The second proposal for presenting *PR*s integrate the distribution of *PR*s besides *mwPR*s. Bornmann and Marx (2014a, 2014b) and more recently Bornmann and Haunschild (2018) proposed to visualize percentiles using beamplots (Doane & Tracy, 2000). The proposal has been taken up by Adams, McVeigh, Pendlebury, and Szomszor (2019) – members of the Institute for Scientific Information (ISI) which is part of Clarivate Analytics: the authors recommend to use beamplots instead of single numbers aggregating percentile distributions (such as in Figure 4). It is an advantage of beamplots that not only annual distributions of percentiles can be presented, but also summary statistics (annual and overall medians). However, it should be considered in the use of beamplots that they are especially suitable for small publication sets (i.e., publication sets of single researchers), since beamplots become unreadable for large sets with many publications.

Figure 5 shows the beamplot of *CP-EX*s for articles published between 2000 and 2005 by Libyan authors. Libya has been selected for this analysis, since there are only 294 articles in the publication set of this study from this country. This number is in that range of paper numbers published by single authors. In the figure, the *PR* of every individual article is visualized using grey diamonds; the annual *mwPR*s (by considering two weights: number of subject categories and number of countries; see sections 4.2 and 4.3) are displayed with increased diamonds (that are black). The vertical black line in Figure 5 shows the *mwPR*s (*CP-EX* = 28.63) across all articles published between 2000 and 2005 by Libyan authors (by considering the two weights mentioned above). The grey dashed line in the figure marks the value 50 – the expected value for an 'average' citation level.

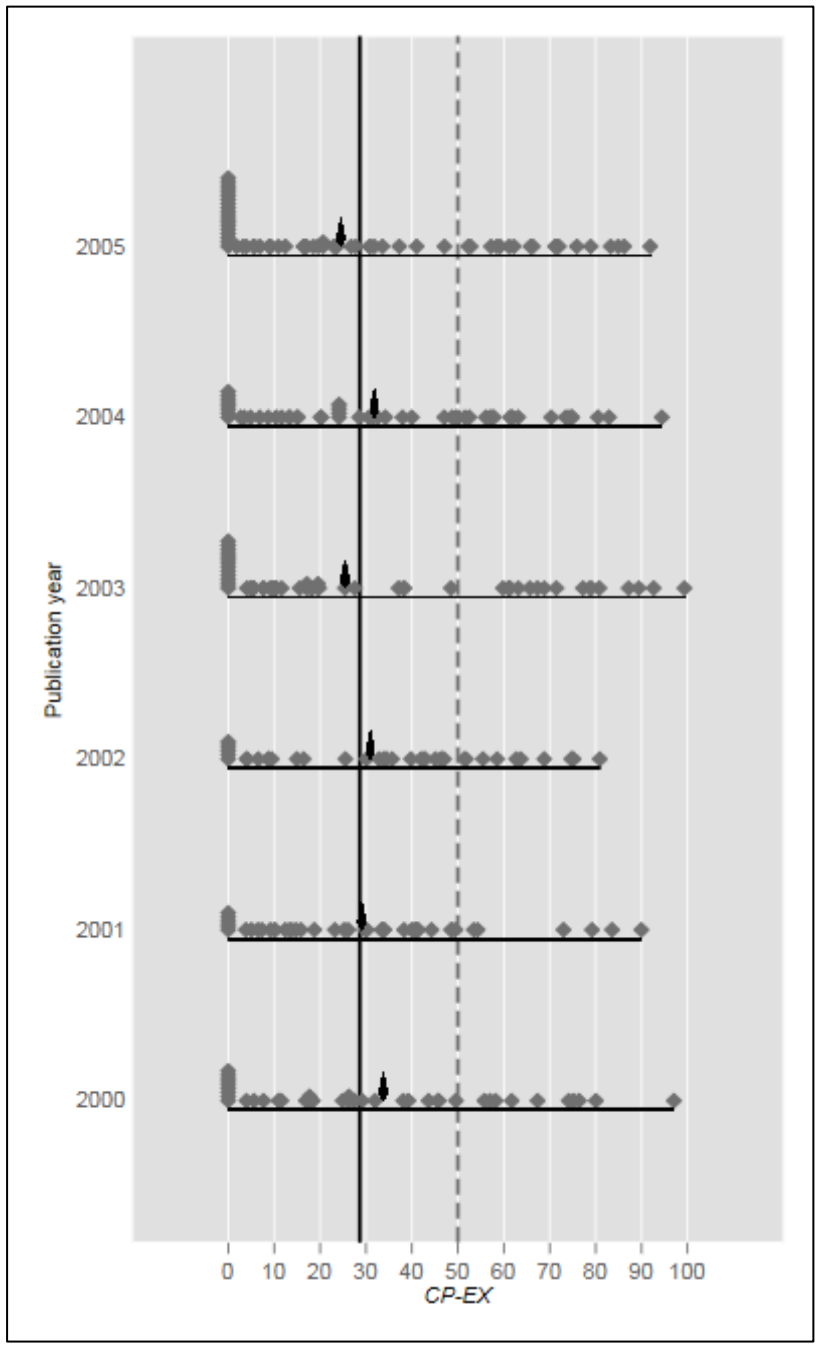

Figure 5. Percentile ranks (*PR*s) beamplot (based on *CP-EX*s) for articles published between 2000 and 2005 by Libyan authors

As the *mwPRs* demonstrates, Libyan authors achieved a citation impact which is significantly below 50: on average, only around 29% of the articles published between 2000 and 2005 in the corresponding subject categories and publication years received a citation impact which is below the impact of the Libyan authors' articles. The distributions of the *PR*s in all publication years demonstrate that the *PR*s are especially concentrated in the low citation impact area (below the 20th *PR*). Highly cited papers – papers which belong to the 10% most frequently cited articles – exist, but they are rare.

As Figure 4 and Figure 5 demonstrate, percentiles can be visualized very differently. Further possibilities of presenting and statistically analyzing percentiles can be found in Bornmann (2013) and Williams and Bornmann (2014).

## 6 Discussion

In a recent study on landmark publications in scientometrics, Tahamtan and Bornmann (2018) worked out that the first method used for time- and field-normalizing citation data was based on percentiles. The introduction of the percentile method to bibliometrics is associated with the name Francis Narin (retired president of CHI Research Inc.). Already at the beginning of the 1980s, McAllister, Narin, and Corrigan (1983) explained citation percentiles as follows: "the *p*th percentile of a distribution is defined as the number of citations $X_p$ such that the percent of papers receiving $X_p$ or fewer citations is equal to *p*. Since citation distributions are discrete, the *p*th percentile is defined only for certain *p* that occur in the particular distribution of interest" (p. 207). Evered, Hamett, and Narin (1989) used the percentage of papers belonging to the 10% most frequently cited papers (named as 'top decile citation performance') to evaluate the citation impact of various institutional units. About 30 years later, Hicks, Wouters, Waltman, de Rijcke, and Rafols (2015) published ten principles to guide research evaluation (using bibliometric data) in *Nature*. According to these authors, "normalized indicators are required, and the most robust normalization method is based on percentiles: each paper is weighted on the basis of the percentile to which it belongs in the citation distribution of its field (the top 1%, 10% or 20%, for example)" (p. 430).

In this study, various approaches have been presented for using percentiles in research evaluation. A very popular approach today is to present the percentage of papers for a unit (e.g., an institution) which belong to the 10% most frequently cited papers: *PP*(top 10%) (see

Bornmann, 2014). This indicator comes under the family of *I3* indicators which combine the number of papers in different *PC*s with specific weights. *P*(top 10%) counts the number of papers belonging to the *PC*s of the 10% most frequently cited papers with a weight of 1. Other *I3* indicators have used up to six *PC*s to measure the citation performance of units with various weights. In section 3.1, it has been argued that *I3* has the disadvantage that information of citation distributions is lost when the data are grouped into (*PR*) classes. Approaches which consider the complete distribution of data should be preferred. Bornmann, Leydesdorff, and Mutz (2013) proposed to use *PP*s which apply the rule by Hazen (1914) to consider the complete citation distributions. The problem with this approach is, however, that *PP*s cannot always be (exactly) interpreted as the percentage of papers (at or) below a certain *CC* (especially when the *PP*s are calculated based on only a few papers).

In recent years, some other percentile approaches have been introduced based on size-frequency distributions with varying advantages and disadvantages – as outlined in section 3.3. In this study, two further approaches (*CP-IN* and *CP-EX*) are introduced which are oriented towards the usual percentile rank definition: *PR x* is defined as the *CC* (at or) below which *x*% of the papers in the combination of publication year and subject category falls. Both approaches can be used very flexible by computing (1) *PR*s for observed *CC*s in distributions and (2) estimated *CC*s for pre-defined *PR*s (e.g., the 90$^{th}$ *PR*). It is one problem for the use of *PR*s in citation analyses that papers in databases such as WoS are frequently assigned to more than one subject category. This problem has been solved by *wPR*s with the consideration of the number of papers in corresponding subject categories. Other problems with *PR*s concern their aggregation: how should *PR*s for papers of various units (e.g., institutions) be aggregated? In sections 4.2 and 4.3, it has been proposed to use *mwPR* and *mwPR(F)*s whereby weights are applied to consider that a paper has been assigned to more than one subject category and/or has been published by more than one unit (e.g., more than one country).

Section 5 addresses the presentation of *PR*s: bar graphs and beamplots can be used to present the results for various units. These graph types are only two examples and other options for visualizing results based on *PR*s exist. Some other options can be found in Bornmann (2013) and Williams and Bornmann (2014). Since the percentile approach has significant advantages over other time- and field-normalizing approaches especially those which are based on mean citation rates, it would be desirable that the percentile approach is more frequently used in bibliometric studies. Although *I3* indicators such as the popular *PP*(top 10%) are robust indicators compared to other time- and field-normalized indicators, they have the important disadvantage that information is lost from the citation distribution. Since *CP-IN* and *CP-EX* consider the complete citation distribution, they should be preferred in bibliometric studies.

# Acknowledgements

The bibliometric data used in this paper are from an in-house database developed and maintained by the Max Planck Digital Library (MPDL, Munich) and derived from the Science Citation Index Expanded (SCI-E), Social Sciences Citation Index (SSCI), Arts and Humanities Citation Index (AHCI) prepared by Clarivate Analytics (Philadelphia, Pennsylvania, USA). We would like to thank Paul Barrett for his support in writing this manuscript and Alexander Tekles and Robin Haunschild for very useful suggestions to improve earlier versions of the manuscript. We are grateful to the reviewer's excellent comments with the objective of improving the paper.